\documentclass[11pt]{JHEP3}
\usepackage{epsfig,multicol,bbm}
\usepackage{cite}
\usepackage{amsmath,amsfonts,amssymb,longtable,mathtools}
\usepackage{array}
\usepackage{graphicx}
\usepackage{subfig}
\usepackage{enumerate}
\usepackage{comment}
\usepackage{bm}
\input epsf.sty

\def\be{\begin{equation}}
\def\ee{\end{equation}}
\def\ba{\begin{array}}
\def\bacc{\begin{array} {cc}}
\def\ea{\end{array}}
\def\bea{\begin{eqnarray}}
\def\eea{\end{eqnarray}}
\def\bd{\begin{displaymath}}
\def\ed{\end{displaymath}}
\def\calN{{\mathcal N}}

\def\Z{\mathbb Z}

\def\L{{\rm L}}

\def\i{{\rm i}}
\def\mod{{\,\,\rm mod}\,\,}

\usepackage{cancel, color}

\title{Discrete ${\bm R}$-Symmetries and Anomaly Universality in Heterotic Orbifolds
}

\author{Nana G.~Cabo Bizet$^{a}$, Tatsuo Kobayashi$^b$, Dami\'an K.~Mayorga Pe\~na$^c$,   \hskip2cm Susha L.~Parameswaran$^d$, Matthias Schmitz$^{c}$, Ivonne Zavala$^e$ \\ \\
${}^a$Centro de Aplicaciones Tecnol\'ogicas y Desarrollo Nuclear, Calle 30, esq.a 5ta Ave, Miramar, 6122 La Habana, Cuba\\
${}^b$Department of Physics, Kyoto University, Kyoto 606-8502, Japan \\
${}^c$Bethe Center for Theoretical Physics and
Physikalisches Institut der Universit\"at Bonn,
Nussallee 12, 53115 Bonn, Germany\\
${}^d$Department of Mathematics and Physics, Leibniz Universit\"at Hannover, Welfengarten 1, 30167 Hannover, Germany\\
${}^e$Centre for Theoretical Physics, University of Groningen,
Nijenborgh 4, 9747 AG Groningen, The Netherlands\\
\\
E-mail: {\email{nana@ceaden.edu.cu}\,,
\email{kobayash@gauge.scphys.kyoto-u.ac.jp}\,,
\email{damian@th.physik.uni-bonn.de}\,,
\email{susha.parameswaran@itp.uni-hannover.de}\,,
\email{mschmitz@th.physik.uni-bonn.de}\,,
\email{e.i.zavala@rug.nl}}
}

\preprint{KUNS-2460 \\ ITP-UH-15/13}

\abstract{
We study discrete $R$-symmetries, which appear in the 4D low energy effective field theory derived from heterotic orbifold models.  We derive the $R$-symmetries directly from the geometrical symmetries of the orbifolds.  In particular, we obtain the corresponding $R$-charges by requiring that the couplings be invariant under these symmetries. This allows for a more general treatment than the explicit computations of correlation functions made previously by the authors, including models with discrete Wilson lines, and orbifold symmetries beyond plane-by-plane rotational invariance.
The $R$-charges obtained in this manner differ from those derived in earlier explicit computations. We study the anomalies associated with these $R$-symmetries, and comment on the results. }


\begin{document}

\newcommand{\zed}{$\mathbb{Z}_2$}

\section{Introduction}\label{introduction}

Discrete symmetries are often imposed in the context of particle physics model building beyond the Standard Model in order to forbid unwanted terms in the Lagrangian.  For example, they have been invoked in order to guarantee the absence of certain operators leading to exceedingly fast proton decay. They have also been very useful for flavor physics to generate textures of quark and lepton masses and mixings.
From a stringy perspective, discrete symmetries are expected to appear, either as discrete remnants of a broken gauge symmetry \cite{gaugediscrete} or to be an inherent property of the compactification from ten to four dimensions.

In this respect, heterotic orbifold compactifications \cite{DHVW} provide a phenomenologically promising UV-complete framework \cite{KRZ,Z6,Z8,Z12,Z2xZ2} that is rich in discrete symmetries, which moreover have intuitive geometric interpretations. One of the features which makes them appealing for phenomenology is the presence of $R$-symmetries.
These $R$-symmetries can be understood as elements of the Lorentz group in the compact space since they treat 4D bosons and
fermions in a different manner. Therefore they are expected not to commute with the generator of the 4D $\mathcal{N}=1$ SUSY algebra. For the specific case of orbifold compactifications, one expects those rotations in $SO(6)$ which are symmetries of the orbifold to manifest as $R$-symmetries of the low energy effective field theory (LEEFT).

The  identification of $R$-symmetries in heterotic models appeared first  in \cite{naturalness} for the  $\mathbb{Z}_3$ orbifold, where they were associated to \emph{orbifold isometries} which in this case were the twists acting on a single plane. Later on, more general expressions were found in \cite{KRZ}.
However, it was recently pointed out by the authors \cite{CKMPSZ}, that in general, the $R$-charges defined in \cite{KRZ}, receive a non-trivial contribution from the so-called {\em gamma-phases}.
In the same way as gauge invariance and other selection rules have been derived \cite{DHVW,HV,DFMS}, the analysis in \cite{CKMPSZ}  was made via the
explicit calculation of vanishing correlation functions, in the absence of discrete Wilson lines.
A classification of the symmetries observed in orbifold geometries  was
presented in the appendix of \cite{CKMPSZ}. There it was observed that such
orbifold symmetries are exhibited by the classical instanton solutions. Thus
one expects them to induce $R$-symmetries in the low energy. In the present note we determine the form of the expected $R$-charges of the physical states, by assuming that the orbifold isometries are also manifest symmetries of the LEEFT.
This allows us to include discrete Wilson lines in the analysis, as well as to consider more general cases. The $R$-charges obtained in this way turn out to differ in the sign of the gamma-phase contribution, with respect to the former derivation.
We provide some possible interpretations of these two results.

Anomalies of discrete symmetries have important implications in 4D field theory \cite{Ibanez:1991hv,ibanez}, in particular string-derived LEEFTs. Such anomalies are expected to cancel via the Green Schwarz mechanism \cite{Green:1984sg}. In heterotic orbifold models, there is only one axion available and hence, all anomalies must be universal, i.e. to cancel up to a common axion shift\footnote{See \cite{Kobayashi:1996pb} and references therein for such universality conditions for $U(1)$ gauge anomalies.}.
In \cite{Araki:2007ss}, anomalies were studied for the discrete $R$-charges corresponding to \cite{KRZ}. Here, we study explicitly the anomaly conditions for the two types of
$R$-charges: the one derived in this paper and the one obtained in \cite{CKMPSZ}. We determine their universality for several orbifolds, factorizable and non-factorizable.  While
anomaly universality could be expected for the $R$-symmetries, there exists no proof that this should be indeed the case.  We discuss this issue and its implications.

The paper is organized as follows: in section \ref{CFT} we review briefly the orbifold CFT; in section \ref{S:RfromS}  we present a derivation of $R$-symmetries from orbifold isometries, which can be applied to the more general cases in which discrete Wilson lines are present. In section \ref{S:Anoml} we compute the anomalies for the $R$-symmetries in several explicit factorizable and non-factorizable examples.
In section \ref{S:Further} we study further possible $R$-symmetries which appear in the low energy effective field theory due to orbifold symmetries that do not leave the fixed points invariant.  We discuss these symmetries, as well as the more familiar $R$-symmetries of section 3, by using an explicit model in \ref{S:Z4}.
In section \ref{S:Discussion} we conclude with the discussion of our results.

\smallskip

{\bf Note added:} While this paper was in preparation, \cite{NRRV} appeared on the arXiv. There the $R$-symmetries of the $\Z_{6II}$ orbifold are derived and the contribution of the discrete Wilson lines is considered for the first time. The $R$-charges derived there also differ in the sign of the gamma phase contribution compared to our previous result \cite{CKMPSZ}. The results presented here were obtained independently and hence confirm (and extend) those in \cite{NRRV}.

\section{String Orbifold CFT}\label{CFT}

We begin by briefly describing the relevant elements of the string orbifold conformal field theory.  Our focus will be on symmetric, six-dimensional orbifolds constructed by modding out a non-freely acting, Abelian isometry of the torus
\[\mathbbm{O}^6=\frac{\mathbbm{T}^6}{P}=\frac{\mathbbm{C}^3/\Lambda}{P}\,=\frac{\mathbbm{C}^3}{S}\,\mathrm{,}\]
where $P$ is the point group and $S=P\ltimes\Lambda$ is the space group. In order to retain $\mathcal{N}=1$ supersymmetry in four dimensions, the point group must be $\mathbbm{Z}_N$ or $\mathbbm{Z}_N\times\mathbbm{Z}_M$. Here we discuss orbifolds in the first class, but our results can easily be  generalized to models of the second type. For $\mathbbm{Z}_N$ orbifolds, the generator of the point group $\theta$ can be brought to the diagonal form
\begin{equation}
\theta=\text{diag}(e^{2\pi{\rm i}v^1},e^{2\pi{\rm i}v^2},e^{2\pi{\rm i}v^3})\, ,
\end{equation}
where the coordinates of the internal space have been taken in a complexified basis $X^i$, $\bar{X}^i$ ($i=1,2,3$).  The twist vector $v=(v^1,v^2,v^3)$ is constrained by $\mathcal{N}=1$ supersymmetry to satisfy $v^1+v^2+v^3=0\,\text{mod}\,1$.

In the orbifold background,  strings can close up to twist and lattice identifications. The closed string boundary conditions involving only lattice identifications give rise to the \emph{untwisted sector}, where the relevant conformal primary fields are the identity operator; the left-moving oscillator fields $\partial X^i$, $\partial \bar{X}^{i}$ (plus their complex
conjugates); the sixteen extra left-movers contributing the Cartans $\partial X^I$ ($I=1,..,16$) and the roots $e^{{\rm i} p \cdot X}$ of the $E_8\times E_8$ gauge symmetry; and the exponential $e^{{\rm i} q^{(a)}\cdot H}$, with
$H$ a five-dimensional vector of free fields corresponding to the bosonized right-moving fermions and $q^{(a)}$ is either a bosonic ($a=1$) of a fermionic ($a=1/2$) weight of $SO(10)$, known as \emph{H-momentum}.

The  boundary conditions of the internal space bosonic coordinates for twisted strings are of the form 
\begin{equation}
 X^i(e^{2\pi {\rm i}}z,e^{-2\pi {\rm i}}\bar{z})=(gX)^i(z,\bar{z})=(\theta^k X)^i(z,\bar{z})+\lambda^i\,,
\label{mono}
\end{equation}
 for any \emph{constructing element} $g=(\theta^k,\lambda)\in S$, where  $z, \bar z$ are the complexified worldsheet coordinates and $\lambda^i\in \Lambda$.  These boundary conditions correspond to strings in the
\emph{$k$-th twisted sector}.  Notice that strings closed by $g$ and $\left\{hgh^{-1} |h \in S \right\}$ are physically equivalent, that is, physical twisted states are associated with conjugacy classes and not space group elements.

The standard way to deal with the branch singularities in \eqref{mono} is to introduce \emph{twist fields} $\sigma(z,\bar{z})$ \cite{HV,DFMS} which serve to implement the local monodromy conditions,
\begin{equation}
\begin{aligned}
&& \partial X^i(z,\bar{z})&\sigma(w,\bar{w})=(z-w)^{-\bar{k}^i}\tau+\ldots\,, \\
 && \partial \bar X^i(z,\bar{z})&\sigma(w,\bar{w})=(z-w)^{-k^i}\tilde{\tau}+\ldots\,,
\end{aligned}
\end{equation}
where  $k^i=kv^i\,\text{mod}\,1$ and $\bar{k}^i=(1-k^i)\,\text{mod}\,1$, such that $0\leq k^i,\bar{k}^i<1$, and $\tau,\,\tilde{\tau}$ are excited twist fields.
The conformal dimensions of $\sigma$ are given by
\begin{equation}
\Delta_\sigma=\bar{\Delta}_\sigma=\frac{1}{2}\sum_{i=1}^3k^i(1-k^i)\, . \label{cw}
\end{equation}
The twist fields for worldsheet fermions can be written in terms of the bosonized fermions as $e^{\i k v \cdot H}$.  This leads to the definition of a shifted H-momentum:
\begin{equation}
q_{sh}^{(a)}=q^{(a)}+k\cdot (0,0,v^1,v^2,v^3)\,,
\end{equation}
so that the primaries in the vertex operators take the familiar form $e^{{\rm i}q^{(a)}_{sh}\cdot H}$.
Furthermore, modular invariance requires the twist to be embedded in the gauge degrees of freedom. In fact, not only the twist but the full space group can be embedded as a shift:
\begin{align}
g=(\theta^k,n_\alpha e_\alpha) \mapsto V_g=kV+n_\alpha W_\alpha\, ,
\end{align}
where $\left\{e_\alpha\right\}$, $\alpha=1,...,6$ spans $\Lambda$, $V$ is the embedding of $\theta$ and the \textit{discrete Wilson lines} $W_\alpha$ are related to the lattice shifts \cite{embed,Kobayashi:1990fx,Kobayashi:1991rp}.   Note that both the group laws of $S$ as well as modular invariance impose several non-trivial constraints on the choice for the embedding vectors, which are summarized for example in \cite{mirage}. The relevant primary for the twisted vertices is
$e^{2\pi {\rm i}p_{sh} \cdot X}$, with $p_{sh}=p+V_g$.

To summarize, the vertex operators describing the emission of twisted states are given by:
\begin{equation}
V_{-a}=e^{-a\phi}\left(\prod_{i=1}^3(\partial X^i)^{\calN_{L}^i}(\partial \bar{X}^{i})^{\bar{\calN}_{L}^{i}}\right) e^{{\rm i}q_{sh}^{(a)}\cdot H}e^
{{\rm i}p_{sh}\cdot X}\sigma\,,
\label{boson}
\end{equation}
where $\phi$ is the superconformal ghost
and the integers $\calN_{L}^i$ and $\bar{\calN}_{L}^{i}$ count,
respectively, the number of left-moving holomorphic and anti-holomorphic oscillators present in the state. Untwisted vertex operators have the same form as those presented before but with all momenta unshifted and with the twist field replaced by the identity operator.  Note that in writing (\ref{boson}), we have taken the four-dimensional momentum to zero, and neglected cocycle\cite{olive} and normalization factors, which are unimportant for our purposes.

Furthermore, invariance of the vertex operators under the full space group needs to be satisfied.
Considering a certain space group element $h$, the bosonic fields transform according to
\begin{equation}
 \partial X^i\xrightarrow{h}e^{2\pi\i v_h^i} \partial X^i\,,\quad X^I\xrightarrow{h}X^I + V_h^I\,,\quad H^i \rightarrow H^i - v_h^i\,.
\end{equation}
In order to see how $h$ acts on the twist fields it is convenient to decompose them into a sum of auxiliary twists $\sigma_g$, one for each element in the conjugacy class $[g]$
\begin{equation}
\sigma\sim\sum_{g^\prime\in[g]}e^{2\pi {\rm i}\tilde{\gamma}(g^\prime)}\sigma_{g^\prime}\,,
\label{sigma}
\end{equation}
where the phases $\tilde{\gamma}(g^\prime)$ will be determined presently.
 For the auxiliary twists one has the following transformation behavior
\begin{equation}
\sigma_{g}\xrightarrow{h}e^{2\pi {\rm i}\Phi(g,h)}\sigma_{hgh^{-1}}\,,
\label{spgt1}
\end{equation}
with the \emph{vacuum phase} $\Phi(g,h)=-\frac{1}{2}\left(V_g\cdot V_h-v_g\cdot v_h\right)$ \cite{mirage}.
This leads to
\begin{equation}
\sigma\xrightarrow{h} e^{2\pi {\rm i}[\gamma_h+\Phi(g,h)]}\sigma\,,
\label{sgtrafo}
\end{equation}
where we have defined
\begin{equation}
\gamma_h=\left[\tilde{\gamma}(g^\prime)-\tilde{\gamma}(hg^\prime h^{-1})\right]\text{ mod } 1,\text{ for some } g^\prime\in[g]\,,
\label{gamma}
\end{equation}
together with the condition
\begin{equation}
-\tilde{\gamma}(hg^\prime h^{-1})+\tilde{\gamma}(g^\prime)=-\tilde{\gamma}(hgh^{-1})+\tilde{\gamma}(g)\mod 1\,,
\label{c1}
\end{equation}
for any pair of elements $g^\prime, g\in [g]$.
Now we can finally see how the vertex \eqref{boson} transforms under
$h$,
\begin{equation}
V_{-a}\xrightarrow{h}\exp\{2\pi {\rm i}[p_{sh}\cdot V_h-v_h^i(q_{sh}^{(a)\hspace{1mm}i}-\calN^i_L+\bar{\calN}^i_L)+\gamma_h+\Phi(g,h)]\}\,V_{-a}\, .
\label{trafo}
\end{equation}
From \eqref{gamma} one can see that if there exists an element $g\in[g]$ which commutes with a certain $h$, then $\gamma_h=0\,\text{mod}\, 1$, such that eq. \eqref{trafo} becomes a projection condition
\begin{equation}
p_{sh}\cdot V_h-v_h^i(q_{sh}^{(a)\hspace{1mm}i}-\calN^i_L+\bar{\calN}^i_L)+\Phi(g,h)=0\mod 1\,,
\end{equation}
which are the so-called \emph{orbifold GSO projectors} responsible for an $\mathcal{N}=1$ supersymmetric spectrum. In all other cases, the gamma-phases can be found by demanding the transformation of $V_{-a}$ to be trivial \cite{Kobayashi:1990mc,Kobayashi:1991rp,Casas:1991ac}, i.e.
\begin{equation}
\gamma_h=-p_{sh}\cdot V_h+v_h^i(q_{sh}^{(a)\hspace{1mm}i}-\calN^i_L+\bar{\calN}^i_L)-\Phi(g,h)=\tilde{\gamma}(g) - \tilde{\gamma}(hgh^{-1}) \mod 1\,.
\label{gammaphase}
\end{equation}
In this way, space group invariance fixes all $\tilde{\gamma}(g^\prime)$ except for one, which can be reabsorbed as an overall phase in $\sigma$.
Notice that there is generically more than one physical twist field $\sigma$ for each conjugacy class, given by the different linear combinations of auxiliary twist fields, in which the different gamma-phase coefficients are determined in terms of the other quantum numbers of the physical state.

\section{Discrete ${\bm R}$-Symmetries from Orbifold Isometries}\label{S:RfromS}

In this section we identify discrete $R$-symmetries which
are expected to appear in the low energy effective field theory of a given orbifold compactification, due to symmetries in the orbifold geometry. Examples of such discrete $R$-symmetries were explicitly verified in \cite{CKMPSZ}, by computing correlation functions.  Our present approach will instead  be to assume that symmetries in the orbifold geometry give rise to $R$-symmetries in the effective field theory and -- given this assumption -- infer the corresponding charge conservation laws. This allows us to be more general, including symmetries beyond the plane-by-plane independent twist symmetries, as well as models with Wilson lines.

The absence of a coupling between L chiral superfields $\Phi_\alpha$ ($\alpha=1,...,\L$) in the superpotential of the LEEFT can be deduced from the vanishing of the
tree-level
 $\L$-point correlator $\psi\psi\phi^{\L-3}$. It is easy to see that the vertex operators $V_{-1}$ and $V_{-1/2}$ corresponding to the bosonic and fermionic fields in a left chiral
supermultiplet are related by a shift in their fermionic weights
\begin{equation}
q^{(1)}_{sh}=q^{(1/2)}_{sh}+({\textstyle \pm\frac{1}{2},\pm\frac{1}{2},-\frac{1}{2},-\frac{1}{2},-\frac{1}{2}})\,\mathrm{.}
\label{chiralmultiplet}
\end{equation}
The computation of the tree-level amplitude requires the emission vertices to cancel the background ghost-charge of two on the sphere. Thus it is necessary to shift the ghost picture of some of the
vertex operators according to
\begin{equation}
V_{0}=e^{\phi}T_{\rm F} V_{-1}\, ,
\label{zeropic}
\end{equation}
where $T_{\rm F}$ is the worldsheet supersymmetry current \cite{Friedan:1985ge,choi}
\begin{equation}
T_{\rm F}=\bar{\partial} X^i\bar{\psi}^i+\bar{\partial} \bar{X}^{i}\psi^i\, ,
\label{wssusycurrent}
\end{equation}
with $\psi^j=\exp\{{\rm i}\, q_j\cdot H\}$ and $q^{i}_j=\delta_j^i$. This
picture-changing operation allows one to write bosonic vertices with zero ghost-charge, at the price of introducing the right-moving oscillators $\bar{\partial} X^i$ and $\bar{\partial} \bar{X}^{i}$ and additional H-momentum. The correlator can then be written as
\begin{equation}
\mathcal{F}=\left<V_{-1/2}(z_1,\bar{z}_1)V_{-1/2}(z_2,\bar{z}_2)V_{-1}(z_3,\bar{z}_3)V_0(z_4,\bar{z}_4)\ldots V_0(z_{\L},\bar{z}_{\L})\right>\,,
\label{correlation}
\end{equation}
where each $V_\alpha=V(z_\alpha,\bar{z}_\alpha)$ represents a certain physical state from the massless spectrum. It is possible to infer several selection rules from the explicit
form of the correlator\footnote{For a review on these selection rules we refer to \cite{ivonne}.}. In the following we will make use of the {\it space group selection rule}
 \begin{equation}
\mathbbm{1}\subset\prod_{\alpha=1}^\L [g_\alpha]\,,
\label{space}
\end{equation}
{\it gauge invariance}
\begin{equation}
\sum_{\alpha=1}^\L p_{sh\hspace{1mm}\alpha}^I=0\, , \label{gaugeinv}
\end{equation}
and {\it H-momentum conservation}
\begin{equation}
\sum_{\alpha=1}^{\L}q_{sh\hspace{1mm}\alpha}^{(1)\hspace{1mm}i}=-1-\calN^{i}_{R}\,,
\label{hmom}
\end{equation}
where $\calN^i_{ R}$ counts the number of holomorphic right-moving oscillators in the correlator.
Using the above rules the correlator \eqref{correlation} can be rewritten in the form
\begin{equation}
\mathcal{F}=\left<\prod_{\alpha=1}^{\rm L}\left(\prod_{i=1}^{3}(\partial X^i)^{\calN^i_{{ L}\hspace{1mm}\alpha}}(\partial \bar{X}^{i})^{\bar{\calN}^{i}_{{L}\hspace{1mm}\alpha}}(\bar{\partial} X^{i})^{\calN^{i}_{ R}}\right)\sigma_{\alpha}\right>\,.
\label{factorizedcorr}
\end{equation}

Let us now deduce the coupling selection rules arising from symmetries of the orbifold geometries. A classification of these symmetries was drawn in the appendix of \cite{CKMPSZ}.  In particular, we are interested in rotations of the torus lattice, which leave the fixed-point structure of the orbifold invariant.
This subgroup of automorphisms was called $D$ in \cite{CKMPSZ}.
Denoting these automorphisms by $\varrho$, they satisfy 
\begin{equation}
\theta=\varrho\theta\varrho^{-1}\quad\text{and}\quad \varrho(g)\in[g]\,\forall g\in S\,,
\label{property}
\end{equation}
and in general, take a block diagonal form, i.e.
\begin{equation}
\varrho=\text{diag}(e^{2\pi{\rm i}\xi^1},e^{2\pi{\rm i}\xi^2},e^{2\pi{\rm i}\xi^3})\,.
\end{equation}
By definition, given a $g \in S$, $\varrho (g)$ is conjugate to $g$, and hence there exists a space group element $h_g$ such that
\begin{equation}
\varrho (g)=h_g g h_g^{-1}\, .
\label{varrho}
\end{equation}
Writing $g=(\theta^k,\lambda)$ and $\varrho(g) = (\theta^k, \varrho \lambda)$, $h_g=(\theta^l,\mu)$ can be determined by finding a solution to the equation
\begin{equation}
\mu = (1-\theta^k)^{-1}(\varrho-\theta^l)\lambda\, .
\end{equation}

In analogy to \eqref{spgt1}, the most general transformation behavior for
the auxiliary twist fields under $\rho$ is given by
\begin{equation}
\sigma_{g}\xrightarrow{\varrho}e^{2\pi {\rm i}\Phi_{\varrho}(g)}\sigma_{\varrho(g)}\,,
\label{twistt}
\end{equation}
which, for the physical twist fields described in eq. \eqref{sigma}, implies
\begin{equation}
\sigma\xrightarrow{\varrho} \sum_{g^\prime\in[g]} e^{2\pi {\rm i}[-\tilde{\gamma}(\varrho(g^\prime))+\tilde{\gamma}(g^\prime)+\Phi_\varrho(g^\prime)]}e^{2\pi {\rm i}\tilde{\gamma}(\varrho(g^\prime ))}\sigma_{\varrho(g^\prime)}\, .
\label{it}
\end{equation}
Since $\varrho$ preserves conjugacy classes, the vertex operators have to be invariant up to phases. This means that we have to require the structure of $\sigma$ to be preserved, which is guaranteed if the following condition is satisfied
\begin{equation}
\tilde{\gamma}(g)-\tilde{\gamma}(\varrho(g))+\Phi_\varrho(g)=\tilde{\gamma}(hgh^{-1})-\tilde{\gamma}(\varrho(hgh^{-1}))+\Phi_\varrho(hgh^{-1})\mod 1\, .
\label{const}
\end{equation}
Using $\varrho(hgh^{-1})=\varrho(h)\varrho(g)\varrho(h)^{-1}$ and the definition \eqref{gamma} we find
\begin{equation}
\Phi_\varrho(hgh^{-1})=\Phi_\varrho(g)+\gamma_h-\gamma_{\varrho(h)}\mod 1.
\label{PhiConjClass}
\end{equation}
Note that this equation implies that once we know the phase $\Phi_\varrho(g)$, the phases for all other elements of the conjugacy class are automatically fixed. Note moreover that the phase, $\Phi_\varrho(g)$, acquired by the auxiliary twist field $\sigma_g$ must depend only on the space-group element $g$, whereas the gamma-phases
associated with the physical twist field $\sigma$ depend, via \eqref{gammaphase},
on the quantum numbers of the corresponding state. Therefore, if \eqref{PhiConjClass} is to be fulfilled for all physical states, the vacuum-phases and gamma-phases must independently fulfill
\begin{equation}
\begin{aligned}
\Phi_\varrho(hgh^{-1})-\Phi_\varrho(g)&=0\mod 1\,, \\
\gamma_h-\gamma_{\varrho(h)}&=0\mod 1 \label{Phi-Phi=0}
\end{aligned}
\end{equation}
for all space group elements $g$, $h$. Now, plugging \eqref{varrho} into \eqref{const} and using \eqref{Phi-Phi=0} we find
\begin{equation}
\gamma_{h_g} = \gamma_{h_{g^\prime}} \mod 1\,,
\end{equation}
for all $g^\prime\in [g]$. This permits the transformation of the $\sigma$ twist to be recast to the desired form
\begin{equation}%
\sigma\xrightarrow{\varrho}  e^{2\pi {\rm i}[\gamma_{h_g}+\Phi_\varrho(g)]}\sigma\, . \label{SigmaUnderRho}
\end{equation}
Note that we have left the phases $\Phi_\varrho(g)$ undetermined.  Finally,  the transformation behavior of the correlator \eqref{factorizedcorr} under $\varrho$ can be concluded. It follows that it is only invariant in the case
\begin{equation}
\sum_{i}\xi^i\left(\sum_{\alpha=1}^\L\left(\calN_{{ L}\hspace{1mm}\alpha}^i-\bar{\calN}_{ L\hspace{1mm}\alpha}^{i}+\calN_{ R\hspace{1mm}\alpha}^{i}\right)\right)+\sum_{\alpha=1}^\L\left(\gamma_{h_{g_\alpha}}+\Phi_{\varrho}(g_\alpha)\right)=0\mod 1\, .
\end{equation}
The phases $\Phi_\varrho$ can be removed from the previous equation, since the space group selection rule together with the OPEs for the twist fields
imply \footnote{Using the space group selection rule, the leading term in the OPE of all auxiliary twist fields involved in the coupling is proportional to the identity, which transforms trivially under $\varrho$.}
\begin{equation}
\sum_{\alpha=1}^\L\Phi_{\varrho}(g_\alpha)=0\mod 1\,.
\end{equation}
The invariance condition for the correlator can now be written in terms of well known quantities when combined with H-momentum conservation \eqref{hmom} and reads
\begin{equation}
\sum_{\alpha=1}^\L\left(\sum_{i=1}^3\xi^i\left[q^{(1)\hspace{1mm}i}_{sh\hspace{1mm}\alpha}-\calN_{{ L}\hspace{1mm}\alpha}^i+\bar{\calN}_{{ L}\hspace{1mm}\alpha}^i\right]-\gamma_{h_{g_\alpha}}\right)=-\sum_{i=1}^3\xi^i\mod 1\, .
\label{rcharge}
\end{equation}
In the case $\sum_{i}\xi^i\neq 0\,\text{mod}\,1$, this condition looks precisely like the coupling selection rule originating from an $R$-symmetry.
In this case, take $M$ to be the smallest integer such that
\begin{equation}
R\equiv -M\sum_{i}\xi^i
\end{equation}
is an integer. Then eq. \eqref{rcharge} takes the more familiar form
\begin{equation}
\sum_{\alpha=1}^\L r_\alpha=R\mod M\,,\,\qquad\text{with}\qquad r_\alpha= \sum_{i=1}^3 M\xi^i\left[q^{(1)\hspace{1mm}i}_{sh\hspace{1mm}\alpha}-\calN_{{ L}\hspace{1mm}\alpha}^i+\bar{\calN}_{{ L}\hspace{1mm}\alpha}^i\right]
		- M\gamma_{h_{g_\alpha}} .
\label{rsymm}
\end{equation}
Thus, by imposing the symmetry of the orbifold generated by $\varrho\in D$ on the correlation function, we have derived a quantity that can be readily interpreted  as a $\mathbb{Z}_M^R$ discrete symmetry of the low energy effective field theory in which $R$ denotes the charge of the superpotential and $r_\alpha$ are the charges of the fields\footnote{For a comprehensive summary of $R$-charge conventions we refer to \cite{Ludeling}.}.

Surprisingly, the discrete symmetry defined in (\ref{rsymm}) does not coincide with the explicit \hbox{$R$-symmetry} result derived in \cite{CKMPSZ}, due to the sign in the last term, which gives the contribution of the gamma-phase to the $R$-charges.  In the following section we discuss this discrepancy in terms of the anomalies for both results.

\section{Universal ${\bm R}$-Symmetry Anomalies}\label{S:Anoml}

In this section we compute the anomalies for the $\mathbb{Z}^R_M$-symmetry derived in the previous section and compare them with the result for the $\mathbb{Z}_M^{R'}$-symmetry derived in \cite{CKMPSZ}. In heterotic orbifold compactifications, there is only one axion available to cancel would-be anomalies via the Green-Schwarz mechanism, so that one typically expects anomalies to be universal. 
Exceptions to this are the anomalies of discrete target-space modular symmetries,
which in many cases can be made universal only after including contributions from one-loop threshold corrections \cite{IL,ferrara}.

For anomalies involving $U(1)$ factors the universality holds up to Ka\v{c}-Moody levels, so we focus only on gravitational and non-Abelian anomalies for which the levels are all equal to 1. Under a $\mathbb{Z}^R_M$ transformation the path integral measure transforms as
\begin{align}
\mathcal{D}\psi\mathcal{D}\overline{\psi}\rightarrow \mathcal{D}\psi\mathcal{D}\overline{\psi}&\exp\left[-2\pi{\rm i} \, \frac{1}{M}\left(\sum_a A_{G_a^2-\mathbb{Z}^R_M}\cdot\frac{1}{16\pi^2}\int \text{tr}\{\mathcal{F}_a\wedge \mathcal{F}_a\}\right.\right.\nonumber\\
&\left.\left.\hspace{3cm}+A_{\text{grav.}^2-\mathbb{Z}^R_M}\cdot\frac{1}{284\pi^2}\int \text{tr}\{\mathcal{R}\wedge \mathcal{R}\} \right)\right]\,,
\end{align}
as can be seen from applying Fujikawa's method \cite{Fujikawa:1979ay}, where the Pontryagin indices
\begin{align}
\frac{T(\mathbf{N}_a)}{16\pi^2}\int \text{tr}\{\mathcal{F}_a\wedge \mathcal{F}_a\}\quad\text{and}\quad\frac{1}{2}\,\frac{1}{284\pi^2}\int \text{tr}\{\mathcal{R}\wedge \mathcal{R}\}
\end{align}
are integer valued \cite{AlvarezGaume:1983ig,AlvarezGaume:1984dr}, and here $T(\mathbf{N}_a)$ denotes the Dynkin index of the fundamental
representation.  The
 corresponding anomaly coefficients are given by
\cite{Ibanez:1991hv,ibanez,Araki:2007ss}\footnote{Recall that gauginos and matter fermions both contribute to the anomaly. The charge of the fermions can be inferred from the piece $\theta\psi\subset \Phi$: if the charge of the multiplet $\Phi$ is denoted by $r$, then the charge of the fermion is $r-R/2$. Analogously, the gauginos appear in the vector multiplet in the form $\overline{\theta}\overline{\theta}\theta\lambda$, so that their charge is $R/2$.}
\begin{align}
A_{G_a^2-\mathbb{Z}^R_M}&=C_2(G_a)\frac{R}{2}+\sum_{\alpha}\left(r_\alpha-\frac{R}{2}\right)T(\mathbf{R}^\alpha_a) \label{nonab}\,,\\
A_{\text{grav.}^2-\mathbb{Z}^R_M}&=\left(-21-1-N_T-N_U+\sum_{a}\text{dim}\{\text{adj}(G_a)\}\right)\frac{R}{2}\nonumber\\
&\hspace{4cm}+\sum_{\alpha}\left(r_\alpha-\frac{R}{2}\right)\cdot\text{dim}\{\mathbf{R}^\alpha\}\label{grav}\,,
\end{align}
with $C_2(G_a)$ being the quadratic Casimir of $G_a$, $\alpha$ running over left chiral matter representations and $T(\mathbf{R}^\alpha_a)$ its corresponding Dynkin index. In eq.~\eqref{grav}, the contributions of $-21$ and $-1$ correspond to the gravitino and dilatino respectively, $N_T$ and $N_U$ are the number of $T$- and $U$-modulini and $a$ runs over all gauge factors (including $U(1)$'s).
If anomalies are cancelled by the same axion shift, given
two gauge factors $G_{a,b}$ the so-called universality conditions must hold
\begin{align}
A_{G_a^2-\mathbb{Z}^R_M}\, \text{mod}\,\, MT(\mathbf{N}_a)&=A_{G_b^2-\mathbb{Z}^R_M}\, \,\text{mod}\,  MT(\mathbf{N}_b)\,,\label{GaugeGaugeUni}\\
A_{G_a^2-\mathbb{Z}^R_M}\,\text{mod}\, \, MT(\mathbf{N}_a)&=\frac{1}{24}\left(A_{\text{grav.}^2-\mathbb{Z}^R_M}\, \,\text{mod}\, \frac{M}{2}\right)\label{GaugeGravUni}\,.
\end{align}
\begin{table}[h!]
\centering
\renewcommand{\arraystretch}{1.4}
\begin{center}
\begin{tabular}{|c|c|c|c|c|c|c|c|c|}
\hline
orbifold & lattice & twist & \multicolumn{3}{c|}{$\varrho$} & $R$ & $M$ \\
\cline{4-6}
 &  &  & $\xi^1$ & $\xi^2$ & $\xi^3$ &  &  \\
\hline
$\mathbb{Z}_4$ & $SO(4)^2\times SU(2)^2$ & $({\textstyle \frac14,\frac14,-\frac24})$ & $1/4$ & $1/4$ & $0$ & $-1$ & $2$ \\
\cline{4-8}
 &  &  & $1/2$ & $0$ & $0$ & $-1$ & $2$ \\
 \cline{4-8}
 &  &  & $0$ & $0$ & $-1/2$ & $+1$ & $2$ \\\hline\hline
$\mathbb{Z}_4$ & $SU(4)^2$ & $({\textstyle \frac14,\frac14,-\frac24})$ & $1/2$ & $0$ & $0$ & $-1$ & $2$ \\
\cline{4-8}
 &  &  & $0$ & $1/2$ & $0$ & $-1$ & $2$ \\\hline\hline
$\mathbb{Z}_{6I}$ & $G_2\times G_2\times SU(3)$ & $({\textstyle \frac16,\frac16,-\frac26})$ & $1/6$ & $1/6$ & $0$ & $-1$ & $3$ \\
\cline{4-8}
 &  &  & $0$ & $0$ & $-1/3$ & $+1$ & $3$ \\\hline\hline
$\mathbb{Z}_{6II}$ & $G_2\times SU(3)\times SU(2)^2$ & $({\textstyle \frac16,\frac26,-\frac36})$ & $1/6$ & $0$ & $0$ & $-1$ & $6$ \\
\cline{4-8}
 &  &  & $0$ & $1/3$ & $0$ & $-1$ & $3$ \\
\cline{4-8}
 &  &  & $0$ & $0$ & $-1/2$ & $+1$ & $2$ \\\hline\hline
$\mathbb{Z}_{8I}$ & $SO(9)\times SO(5)$ & $({\textstyle \frac18,-\frac38,
\frac28})$
& $1/4$ & $-3/4$ & $0$ & $+1$ & $2$ \\
\cline{4-8}
 &  &  & $0$ & $0$ & $1/2$ & $-1$ & $2$ \\\hline\hline
$\mathbb{Z}_{8II}$ & $SO(8)\times SO(4)$ & $({\textstyle \frac18,\frac38,-\frac48})$ & $1/8$ & $3/8$ & $0$ & $-1$ & $2$ \\
\cline{4-8}
 &  &  & $0$ & $0$ & $-1/2$ & $+1$ & $2$ \\\hline\hline
$\mathbb{Z}_{12I}$ & $SU(3)\times F_4$ & $({\textstyle \frac{4}{12},\frac{1}{12},-\frac{5}{12}})$ & $1/3$ & $0$ & $0$ & $-1$ & $3$ \\
\cline{4-8}
 &  &  & $0$ & $1/12$ & $-5/12$ & $+1$ & $3$ \\\hline\hline
$\mathbb{Z}_{12II}$ & $F_4 \times SO(4)$ & $({\textstyle \frac{1}{12},\frac{5}{12},-\frac{6}{12}})$ & $1/12$ & $5/12$ & $0$ & $-1$ & $2$ \\
\cline{4-8}
 &  &  & $0$ & $0$ & $-1/2$ & $+1$ & $2$ \\\hline\hline
\end{tabular}
\end{center}
\caption{Summary of point groups studied with their corresponding lattices and corresponding orbifold isometries. The charge of the superpotential $R$ and the order of the symmetry $M$ are also given.}
\label{tab:T1}
\end{table}

 Let us focus on the orbifolds presented in table \ref{tab:T1}, with
 the isometries discussed in \cite{CKMPSZ}.
As examples, we give the
 space group elements $h_g$ needed to calculate the gamma-phases for
 the $R$-symmetries identified for the $\mathbb{Z}_4$ and
 $\mathbb{Z}_{6II}$ orbifolds in the appendix.
  We used the C++ orbifolder \cite
 {orbifolder} to compute the spectrum and the corresponding anomalies for all of the embeddings classified in \cite{Katsuki:1989kd,Katsuki:1989bf} \emph{without Wilson lines}, with the
$R$-charge assignment given in eq. \eqref{rsymm} for the factorizable $\mathbb{Z}_4$, $\mathbb{Z}_{6I}$ and $\mathbb{Z}_{6II}$, as well as the non-factorizable $\mathbb{Z}_{8I}$ orbifold.
In all models the $R$-anomalies satisfy universality conditions. When taking
the $R$-charges without the gamma contribution \cite{KRZ}, universality is particular to very few models. The same is observed when using the opposite sign for the $\gamma$ phases, as derived in \cite{CKMPSZ}.
Furthermore we considered models \emph{with Wilson lines}. For each of the allowed shift embeddings we randomly generated 10 000 Wilson line configurations and found that in these cases the $R$-charges computed from eq. \eqref{rsymm} show universality relations for all orbifolds studied.
This is an overwhelming result and a strong hint that the $R$-charges derived here are correct. However, the reason for the opposite sign in front of the gamma-phase contribution derived in \cite{CKMPSZ} remains to be understood. Here, we discuss a possible way out. We have assumed the auxiliary twist fields $\sigma_g$ to transform according to \eqref{spgt1}.  Suppose instead that the auxiliary twist fields $\sigma_g$ have the (albeit counter-intuitive) transformation law
\begin{equation}
\sigma_{g}\xrightarrow{h}e^{2\pi {\rm i}\Phi(g,h)}\sigma_{h^{-1}gh}\,,
\label{sigma2}
\end{equation}
where the role of $h$ and $h^{-1}$ is interchanged compared to \eqref{spgt1}. The resulting $R$-charge assignment \eqref{rcharge},\eqref{rsymm} is independent of this change. Indeed, if one goes through the derivations in section 3 using this transformation behavior one arrives at
\begin{align*}
\sigma\xrightarrow{\varrho}  e^{2\pi {\rm i}[\gamma_{h^{-1}_g}+\Phi_\varrho(g)]}\sigma\,
\end{align*}
instead of \eqref{SigmaUnderRho}, while \eqref{gammaphase} gets modified to
\begin{align*}
\gamma_h=p_{sh}\cdot V_h-v_h^i(q_{sh}^{(a)\hspace{1mm}i}-\calN^i_L+\bar{\calN}^i_L)+\Phi(g,h)\,,
\end{align*}
so that the resulting $R$-charge conservation law remains precisely \eqref{rsymm}.  Meanwhile, the same $R$-charge conservation law \eqref{rsymm} would be derived by the explicit computations in \cite{CKMPSZ}.

\section{Further ${\bm R}$-Symmetry Candidates}\label{S:Further}

Having observed universality for the $R$-symmetries derived above, let us now elaborate on an additional set of symmetries which was already introduced in \cite{CKMPSZ}. There we observed that some lattice automorphisms exchange certain fixed points of the same twisted sector. We denoted the subgroup of lattice automorphisms satisfying this property by $F$. At first it seems that this kind of isometry has nothing to do with the string orbifold compactification. In some cases, however, one observes that the fixed points which get mapped
to each other under a certain $\zeta\in F$ allocate identical matter representations, and hence it gives rise to a symmetry in the low energy effective field theory\footnote{Analogous symmetries were considered in \cite{flavour}, but as these were permutation symmetries rather than rotational symmetries, they did not correspond to $R$-symmetries.}.

Now we are concerned with the computation of the charges of the fields under this new type of symmetry.   We will use the fact that, for all cases considered, the elements in $F$ can be written in a block diagonal form as
\begin{equation}
\zeta=\text{diag}(e^{2\pi{\rm i}\eta^1},e^{2\pi{\rm i}\eta^2},e^{2\pi{\rm i}\eta^3}). \label{zeta}
\end{equation}As expected, for those vertex operators which are eigenstates of $\zeta$, the charges are similar to \eqref{rsymm}:
\begin{equation}
r_\alpha= \sum_{i=1}^3 M\eta^i\left[q^{(a)\hspace{1mm}i}_{sh\hspace{1mm}\alpha}-\calN_{{ L}\hspace{1mm}\alpha}^i+\bar{\calN}_{{ L}\hspace{1mm}\alpha}^i\right]
		- M\gamma_{h_{g_\alpha}} .
\end{equation}
 Let us therefore consider vertices of states located at non-invariant fixed points. For simplicity, let us assume that $\zeta$ only interchanges certain conjugacy classes, i.e.
\begin{equation}
 [g]\xleftrightarrow{\zeta}[g^\prime]\,, \quad
 g\nsim g^\prime\, .\label{f}
\end{equation}
This implies that a vertex $V$ from $[g]$ gets mapped to its counterpart $V^\prime$, where $V$ and $V^\prime$ share the same quantum numbers\footnote{Note that although $V$ and $V^\prime$ are associated with different conjugacy classes, the conjugacy class is not a good quantum number to distinguish the states.
 Moreover, note that if the coupling $\langle V_1 \dots V_{\L-1} V \rangle$ is allowed by all selection rules,
then so is $\langle V_1 \dots V_{\L-1} V^\prime\rangle$.}.
Writing
\begin{eqnarray}
\sigma &\sim&\sum_{g \in[g]}e^{2\pi {\rm i}\tilde{\gamma}(g)}\sigma_{g}\,, \nonumber \\
\sigma^\prime &\sim&\sum_{g^\prime\in[g^\prime]}e^{2\pi {\rm i}\tilde{\gamma}^\prime(g^\prime)}\sigma_{g^\prime}\,, \nonumber
\end{eqnarray}
 the twist fields involved in these vertices will then transform according to\footnote{In general one can allow for vacuum phases for the twist fields under this transformation. However they turn out to be irrelevant for our discussion in the same way as already observed in section \ref{S:RfromS}.}
\begin{equation}
\begin{aligned}
\sigma &\xrightarrow{\zeta}\exp\{2\pi {\rm
 i}[\tilde{\gamma}(g)-\tilde{\gamma}^\prime(\zeta(g))+\Phi_\zeta(g)]\}\sigma^\prime\,,\\
\sigma^\prime &\xrightarrow{\zeta}\exp\{2\pi {\rm
 i}[\tilde{\gamma}^\prime(g^\prime)-\tilde{\gamma}(\zeta(g^\prime))+\Phi_\zeta(g^\prime)]\}\sigma\,,
 \end{aligned}
  \label{SigmaUnderZeta}
\end{equation}
and therefore
\begin{equation}
\begin{aligned}
V &\xrightarrow{\zeta}\exp\{2\pi\i[ -\eta^i(q^{(a)\hspace{1mm}i}_{sh}-\calN^i_L+\bar{\calN}^i_L) + \tilde{\gamma}(g)-\tilde{\gamma}^\prime(\zeta(g))+\Phi_\zeta(g)]\}V^\prime\,,\\
V^\prime &\xrightarrow{\zeta}\exp\{2\pi\i[ -\eta^i(q^{(a)\hspace{1mm}i}_{sh}-\calN^i_L+\bar{\calN}^i_L) +\tilde{\gamma}^\prime(g^\prime)-\tilde{\gamma}(\zeta(g^\prime))+\Phi_\zeta(g^\prime)]\}V\,.
 \end{aligned}
  \label{VUnderZeta}
\end{equation}
Recall that $q^{(a)\hspace{1mm}i}_{sh}$, $\calN^i_L$ and $\bar{\calN}^i_L$ are the same for both $V$ and $V^\prime$.
Note that, a priori, the transformation phases in \eqref{SigmaUnderZeta}, \eqref{VUnderZeta} cannot be related to physical gamma-phases since $g$ and $\zeta(g)$ belong to different conjugacy classes.
Since $V$ and $V^\prime$ differ only in their conjugacy classes and carry identical quantum numbers, one expects couplings involving either $V$ and $V^\prime$ to differ only by constant phases.
One can write the vertices $V$ and $V^\prime$ in a basis of eigenstates of $\zeta$
\begin{equation}
 V^{(s)}=V + e^{2\pi {\rm i}(\delta+s)}V^\prime\, ,\quad s=0\,,{\textstyle \frac12}\,, \label{VTilde}
\end{equation}
in which $\delta$ is a phase fixed so that the operators $V^{(s)}$ transform indeed only up to a phase under $\zeta$.
Using equations \eqref{VUnderZeta} and \eqref{VTilde}, we can fix $\delta$ and write the transformation behavior of $V^{(s)}$ under $\zeta$ as
\begin{equation}
V^{(s)}\xrightarrow{\zeta} \exp\left\{ 2\pi {\rm i}\left[-\eta^i(q^{(a)\hspace{1mm}i}_{sh}-\calN^i_L+\bar{\calN}^i_L)+\textstyle{\frac{1}{2}}\left(\gamma_{h_g}+\gamma^\prime_{h_{g^\prime}}\right)+s\right]\right\}
 V^{(s)}\,.\label{Vs}
\end{equation}
Here the space group elements $h_g$ and $h_{g^\prime}$ are defined such that
\begin{equation}
 \zeta(g^\prime)=h_g g h_g^{-1}\, ,\quad
 \zeta(g)=h_{g^\prime} g^\prime h_{g^\prime}^{-1}\,,
\end{equation}
for any combination of representatives $g$ and $g^\prime$ in the same way as described in section \ref{S:RfromS}, and recall that $\gamma_{h_g} = \tilde\gamma(g) - \tilde\gamma(h_g g h_g^{-1})$ and $\gamma^\prime_{h_{g^{\prime}}} = \tilde\gamma^\prime(g^{\prime}) - \tilde\gamma^\prime(h_{g^{\prime}} g^\prime h_{g^{\prime}}^{-1})$.
From the transformation property of the $V^{(s)}$ we can now read off their corresponding $R$-charges
\begin{equation}
r^{(s)}=M\sum_{i=1}^3\eta^i(q^{(a)\hspace{1mm}i}_{sh}-\calN^i_{L}+\bar{\calN}^i_{L})-\textstyle{\frac{1}{2}}M\left(\gamma_{h_g}+\gamma^\prime_{h_{g^\prime}}\right)-Ms\,,
\end{equation}
where $M$ is the smallest integer such that
\begin{equation}
R\equiv -M\sum_{i=1}^3\eta^i\in\mathbb{Z}\,.
\end{equation}
When writing the low energy effective field theory in terms of the $\zeta$-eigenstates, the corresponding R-charge conservation law implies that any coupling must vanish unless
\begin{equation}
\sum_{\alpha=1}^\L r_\alpha=R\mod M\,.
\end{equation}

The result we just obtained for the elements in $F$, at least for the case of factorizable orbifolds, has remarkable implications. $R$-symmetries of the LEEFT are not only due to those remnants of the Lorentz group which leave the fixed points invariant. Even those automorphisms mapping different fixed point conjugacy classes to each other can source $R$-symmetries in the field theory. In contrast to those emerging from symmetries in $D$, these novel $R$-symmetries can be broken by
Wilson line configurations that spoil the degeneracy of the states located at the non-invariant fixed points. Note that in our derivation we assumed that $\zeta$ at most \emph{interchanges} pairs of conjugacy classes, but in principle
more intricate transformation patterns can emerge, particularly in the case of non-factorizable orbifolds. We expect that in those cases, the charges can be computed in a similar fashion.

\section{An Explicit Example: $\boldsymbol{\mathbb{Z}_4}$ on $\boldsymbol{SO(4)^2\times SU(2)^2}$}

\label{S:Z4}

Here we illustrate our results by discussing in detail the $\mathbb{Z}_4$ orbifold on the lattice of $SO(4)^2\times SU(2)^2$ with the twist as given in table \ref{tab:T1}. One easily
sees that a basis of generators for the group $D$ is given by
\begin{align}
&\varrho_1=\theta_1\theta_2=(e^{2\pi {\rm i}\frac14},e^{2\pi {\rm i}\frac14},1)\,,\quad \varrho_2=(\theta_1)^2=(e^{2\pi {\rm i}\frac12},1,1)\,,\nonumber \\
&\hspace{3cm}\varrho_3=\theta_3=(1,1,e^{-2\pi {\rm i}\frac14})\,.
\end{align}
As stressed before, each of these symmetries leads to universal anomalies in all of the models studied. As an example let us discuss the following shift embedding and Wilson line configuration
\begin{equation}
\begin{aligned}
V& =\left({\textstyle -1,  -\frac34,     0,     0,     0,     0,     0,   \frac14,
0,     0,     0,     0,     0,     0,     0,   \frac12}\right)\, ,\\
W_1 =W_2 &= \left({\textstyle  \frac74,   \frac14,  -\frac34,  -\frac14,   \frac14,   \frac14,   \frac54,   \frac14,
-\frac74,  -\frac14,  -\frac14,  -\frac14,  -\frac14,  -\frac14,   \frac14,   \frac74}\right)\,,\\
  W_3 = W_4 &= \left({\textstyle  -\frac12,  -\frac32,  -\frac32,     1,  -\frac32,   \frac32,   \frac32,     1,
-\frac14,  -\frac74,  -\frac54,  -\frac14,   \frac14,   \frac14,   \frac34,  -\frac74}\right)\,,\\
  W_5 &= \left({\textstyle     0,  -\frac12,   \frac32,  -\frac32,   \frac32,   \frac32,     1,   \frac32,
1,    -2,     0,     1,   \frac12,     2,     1,  -\frac32}\right)\,,\\
W_6 &= 0\,.
\end{aligned}
\end{equation}
Recall the identifications for the Wilson lines: $W_1 \sim W_2$, $W_3 \sim W_4$, see figure \ref{Z4Config} where
the Wilson line configuration for this model is shown.
\FIGURE{
\includegraphics[width=12.8cm]{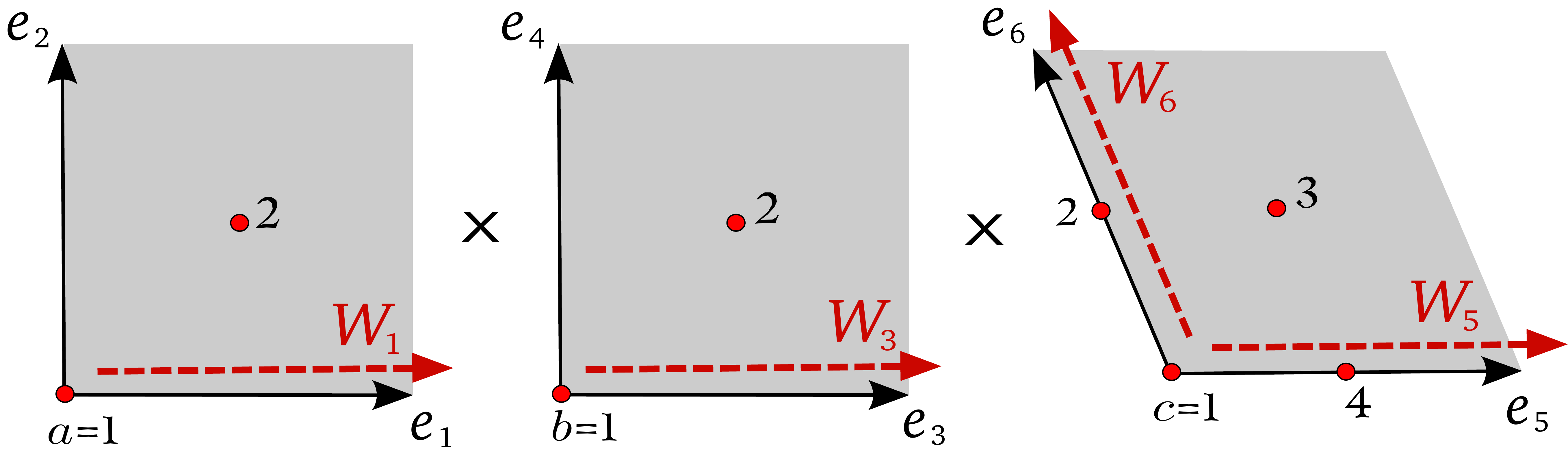}
\caption{Wilson line configuration for the $\mathbb{Z}_4$ orbifold studied in the text.}
\label{Z4Config}}

This embedding leaves the following gauge symmetry unbroken
\[SU(4)_1 \times SU(2)_1 \times SU(2)_2\times SU(4)_2 \times SU(2)_3
\times U(1)^7\subset E_8\times E_8 \, .\]\,
The anomaly coefficients obtained for this specific orbifold model are
\begin{equation}
\begin{aligned}
   A_{\text{grav.}^2 - \varrho_1} &= -76\,, & A_{\text{grav.}^2 - \varrho_2} &= 94\,, &    A_{\text{grav.}^2 - \varrho_3} &= 84\,,\\
   A_{SU(4)_1^2 - \varrho_1} &= -3\,, &     A_{SU(4)_1^2 - \varrho_2} &= 3\,, &     A_{SU(4)_1^2 - \varrho_3} &= -1\,,\\
   A_{SU(2)_1^2 - \varrho_1} &= -5\,, &  A_{SU(2)_1^2 - \varrho_2} &= 1\,,   &  A_{SU(2)_1^2 - \varrho_3} &= 5\,,\\
   A_{SU(2)_2^2 - \varrho_1} &= -11\,, & A_{SU(2)_2^2 - \varrho_2} &= 6\,, & A_{SU(2)_2^2 - \varrho_3} &= 5\,,\\
   A_{SU(4)_2^2 - \varrho_1} &= -3\,,  & A_{SU(4)_2^2 - \varrho_2} &= -1\,,  &A_{SU(4)_2^2 - \varrho_3} &= -1\,,\\
   A_{SU(2)_3^2 - \varrho_1} &= -11\,, & A_{SU(2)_3^2 - \varrho_2} &= 3\,,  & A_{SU(2)_3^2 - \varrho_3} &= 5\,.
   \end{aligned}
\end{equation}
One can straightforwardly check that all of these values satisfy the universality conditions \eqref{GaugeGaugeUni} and \eqref{GaugeGravUni}.

This model also serves to discuss the effects of the new $R$-symmetries emerging from $F$. Note that \begin{equation}
 \zeta=\theta_1=(e^{2\pi {\rm i}\frac14},1,1)\in F\,,
\end{equation}
interchanges the fixed points
\begin{equation}
z_g=\frac{e_2+e_3}{2}\xleftrightarrow{\zeta}z_{g^\prime}=\frac{e_2+e_4}{2}\,,
\end{equation}
of the second twisted sector $T_2$, which are generated by space group elements $g$ and $g^\prime$ from different conjugacy classes.
This is illustrated in figure \ref{IndepTwist}.
\FIGURE{
\includegraphics[width=12.8cm]{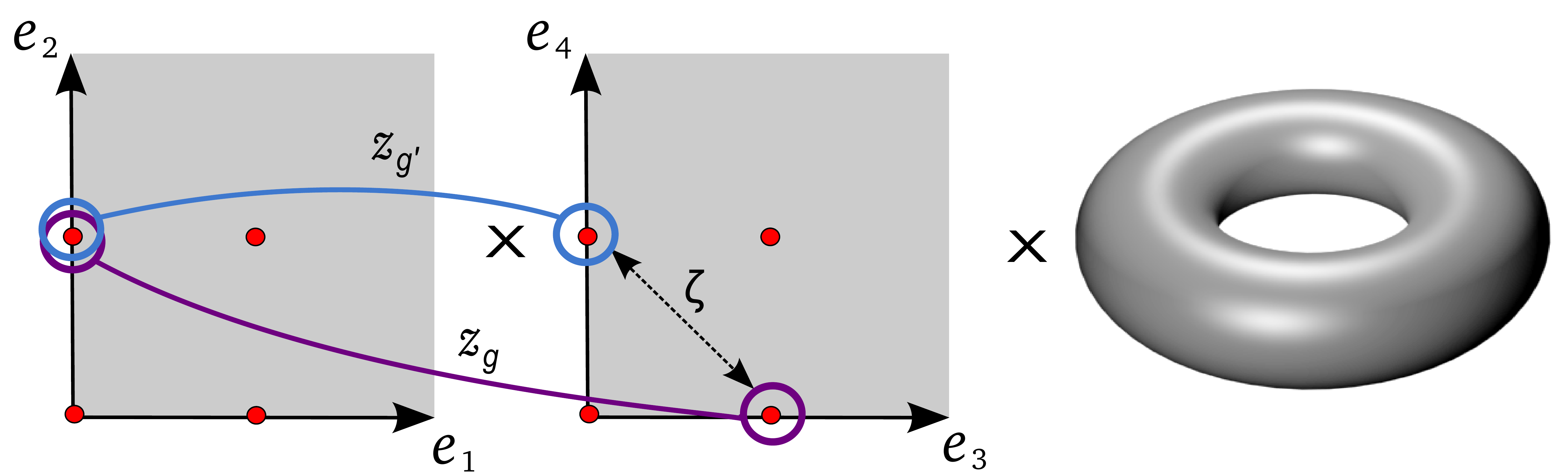}
\caption{Representation of the $\theta_1$ action on the $T_2$ sector fixed points of the $\mathbb{Z}_4$ orbifold studied.}
\label{IndepTwist}}

Note that
as we have $W_1=W_2$, $W_3=W_4$,  the transformation $\zeta$ respects the Wilson line structure. Hence the spectrum contains identical states $V$ and $V^\prime$ sitting at each of
the relevant fixed points. As an example consider the states specified by the following quantum numbers
\begin{align}
p_{sh} &=\left({\textstyle-\frac34, \frac14, -\frac14, -\frac14, -\frac14, -\frac14, -\frac14,
-\frac14, 0, 0, -\frac12, \frac12, 0, 0, 0, 0}\right)\,,\\
q_{sh} &=\left({\textstyle 0, -\frac12, -\frac12, 0}\right)\,,
\end{align}
with no left-moving oscillators. The $p_{sh}$ presented is the highest weight of the representation $(\mathbf{1},\mathbf{2},\mathbf{2},\mathbf{1},\mathbf{2})$ with all $U(1)$ charges equal to zero. Two identical copies of this state live at the fixed points under
consideration. The elements $h_g$ and $h_{g^\prime}$ needed to compute the $R$-charges are given by
\begin{equation}
\label{ExampleHgs}
 h_g=(\theta,e_3)\quad\text{and}\quad  h_{g^\prime}=(\theta,0)\,,
\end{equation}
and the corresponding gamma-phases are $\gamma(g)=\gamma(g^\prime)=3/4$. With this information we can compute the $R$-charges for the eigenstates of $\zeta$ to be
\begin{equation}
 r^{(s)}=-7/2-4s\,.
\end{equation}
We also computed the anomaly coefficients for the $R$-symmetry $\zeta$, with a scan
of over 100.000 randomly generated models. In all cases the anomalies turned out to be universal. Similar results are to be expected for
all orbifolds for which the group $F$ is non-trivial. Note that, in our example, $\zeta^2=\varrho_2$ and one can show that the $R$-charges under $\varrho_2$ are twice those under $\zeta$ up to multiples of 2. This implies that one can safely take $\zeta$ and $\varrho_1$ as a basis for all $R$-symmetries in the factorizable $\mathbb{Z}_4$ orbifold.

\section{Discussion} \label{S:Discussion}

In this work we have derived $R$-symmetries expected in the low energy effective field theory of heterotic orbifold compactifications, directly from the symmetries observed in the orbifold geometry. In particular, by imposing that the string correlation functions are invariant under
such symmetries, we were able to infer the $R$-charges that are conserved in the low energy theory.  This approach allowed us to be more general than the explicit computations of vanishing correlation functions pursued in \cite{CKMPSZ}.  For example, we were presently able to treat orbifold models with discrete Wilson lines.
Moreover, we identified new $R$-symmetries, which arise from rotations which interchange inequivalent fixed points supporting the same physical states.

The conserved $R$-charges associated with 
rotations preserving the fixed point structure of the orbifold were derived in explicit models for all $\Z_N$ orbifold models, with and without discrete Wilson lines.  The corresponding anomalies were then computed, and a scan of thousands of randomly generated models showed that the anomalies were universal. 
This is also the case for the $R$-symmetries conjectured for non-factorizable orbifolds, even though there are some non-trivial steps still missing for the full understanding of their CFT.
Further, we identified an additional source for $R$-symmetries, namely those isometries under which certain fixed points (that support the same twisted matter) get exchanged. An example was given for a $\Z_4$ orbifold. It is remarkable that the corresponding $R$-symmetry anomalies were also found to satisfy universality relations.

The universality of the $R$-symmetry anomalies is certainly a beautiful and compelling result.  On the other hand, the $R$-charges that were obtained from the explicit computation of vanishing string correlation functions \cite{CKMPSZ}, have the opposite sign in the gamma-phase contribution, and do not always lead to the universal anomalies. It remains an essential open question to understand the reason behind this mismatch, although we have pointed out a possible origin for the discrepancy.
Moreover, it seems important to bear in mind the following observations.  Anomaly universality does not necessarily guarantee that a symmetry is an exact symmetry.
Examples in which anomalies are universal, but the symmetry is explicitly broken by non-perturbative effects, are some continuous target-space modular symmetries \cite{IL,Ibanez:1999pw}.
Meanwhile, anomalies which are non-universal might be partially cancelled by one-loop threshold effects, as sometimes observed for discrete target-space modular invariance \cite{IL,ferrara}. Finally, as discrete symmetries are by definition global there is no inconsistency if they happen to be anomalous.  Simply, this would imply that they are not symmetries in the full quantum theory.

Despite the fact that the lattices studied here are the simplest possibilities, we expect similar results for the more general orbifold models discussed in  \cite{Konopka,Fischer:2012qj}.
It remains to be discussed how these redefined $R$-charges affect the phenomenology of MSSM like models found all over the orbifold landscape.
In those models where the top Yukawa coupling is purely untwisted, one can guarantee its survival. However, in order to address issues such as Yukawa textures, decoupling of the exotics and proton decay it is necessary to look at explicit models.
An important question concerns the effects of the new $R$-charge redefinitions in the particular context of $\mathbb{Z}_2\times\mathbb{Z}_2$ \cite{Z2xZ2}, especially in those models where the famous $\mathbb{Z}_4^R$ symmetry of \cite{Lee:2010gv} could be realized. Another interesting issue has to do with the fact that the $R$-charges now receive contributions from the gauge part of the theory, so it is worth studying how the gauge bundle information enters the $R$-symmetries that one also expects to see in the orbifold phase of gauged linear sigma models, as well as in partial blow-ups.

\section*{Acknowledgments}

We would like to thank M.~Blaszczyk, S.~F\"orste, P.~Oehlmann and F.~R\"uhle  for useful discussions.
N.~G.~C.~B.~is supported by ``Proyecto Nacional de Ciencias B\'asicas Particulas y Campos'' (CITMA, Cuba). T.~K.~is supported in part by  the Grant-in-Aid for the Scientific Research No. 25400252 from the Ministry of Education, Culture, Sports, Science and Technology of Japan. S.~L.~P.~is funded by Deutsche Forschungsgemeinschaft inside the ``Graduiertenkolleg GRK 1463''.
The work of  D.~K.~M.~P.~and M.~S.~was partially supported by the SFB-Tansregio TR33
``The Dark Universe" (Deutsche Forschungsgemeinschaft) and the European Union 7th network program ``Unification
in the LHC era" (PITN-GA-2009-237920).

\begin{appendix}

\section{Space group elements $h_g$  for the $\Z_4$ and $Z_{6II}$ orbifolds}

In this appendix we present the values of $h_g$ for $\mathbbm{Z}_4$ and $\mathbbm{Z}_{6II}$ used in the main text, in tables \ref{S:AppHgZ4} and \ref{S:AppHgZ6II}.

\begin{table}[h]
\centering
\small
\begin{tabular}{  c | c  c  c  c }
$g$ & $h_g^{\theta_1\theta_2}$ & $h_g^{\theta_3}$ & $h_g^{\theta_1^2}$ & $h_g^{\theta_1}$\\
\hline
$\mathbbm{1}, (0, 0, 0, 0, 0, 0)$ & $\mathbbm{1}, (0, 0, 0, 0, 0, 0) $ & $\mathbbm{1}, (0, 0, 0, 0, 0, 0)$ & $\mathbbm{1}, (0, 0, 0, 0, 0, 0)$ & $\mathbbm{1}, (0, 0, 0, 0, 0, 0)$\\
$\theta, (0, 0, 0, 0, 0, 0)$ & $\mathbbm{1}, (0, 0, 0, 0, 0, 0) $ & $\mathbbm{1}, (0, 0, 0, 0, 0, 0)$ & $\mathbbm{1}, (0, 0, 0, 0, 0, 0)$ & $\mathbbm{1}, (0, 0, 0, 0, 0, 0)$\\
$\theta, (0, 0, 0, 0, 0, 1)$ & $\mathbbm{1}, (0, 0, 0, 0, 0, 0) $ & $\mathbbm{1}, (0, 0, 0, 0, 0, -1)$ & $\mathbbm{1}, (0, 0, 0, 0, 0, 0)$ & $\mathbbm{1}, (0, 0, 0, 0, 0, 0)$\\
$\theta, (0, 0, 1, 0, 0, 0)$ & $\mathbbm{1}, (0, 0, -1, 0, 0, 0) $ & $\mathbbm{1}, (0, 0, 0, 0, 0, 0)$ & $\mathbbm{1}, (0, 0, 0, 0, 0, 0)$ & $\mathbbm{1}, (0, 0, 0, 0, 0, 0)$\\
$\theta, (0, 0, 1, 0, 0, 1)$ & $\mathbbm{1}, (0, 0, -1, 0, 0, 0) $ & $\mathbbm{1}, (0, 0, 0, 0, 0, -1)$ & $\mathbbm{1}, (0, 0, 0, 0, 0, 0)$ & $\mathbbm{1}, (0, 0, 0, 0, 0, 0)$\\
$\theta, (1, 0, 0, 0, 0, 0)$ & $\mathbbm{1}, (-1, 0, 0, 0, 0, 0) $ & $\mathbbm{1}, (0, 0, 0, 0, 0, 0)$ & $\mathbbm{1}, (-1, -1, 0, 0, 0, 0)$ & $\mathbbm{1}, (-1, 0, 0, 0, 0, 0)$\\
$\theta, (1, 0, 0, 0, 0, 1)$ & $\mathbbm{1}, (-1, 0, 0, 0, 0, 0) $ & $\mathbbm{1}, (0, 0, 0, 0, 0, -1)$ & $\mathbbm{1}, (-1, -1, 0, 0, 0, 0)$ & $\mathbbm{1}, (-1, 0, 0, 0, 0, 0)$\\
$\theta, (1, 0, 1, 0, 0, 0)$ & $\mathbbm{1}, (-1, 0, -1, 0, 0, 0) $ & $\mathbbm{1}, (0, 0, 0, 0, 0, 0)$ & $\mathbbm{1}, (-1, -1, 0, 0, 0, 0)$ & $\mathbbm{1}, (-1, 0, 0, 0, 0, 0)$\\
$\theta, (1, 0, 1, 0, 0, 1)$ & $\mathbbm{1}, (-1, 0, -1, 0, 0, 0) $ & $\mathbbm{1}, (0, 0, 0, 0, 0, -1)$ & $\mathbbm{1}, (-1, -1, 0, 0, 0, 0)$ & $\mathbbm{1}, (-1, 0, 0, 0, 0, 0)$\\
$\theta, (0, 0, 0, 0, 1, 0)$ & $\mathbbm{1}, (0, 0, 0, 0, 0, 0) $ & $\mathbbm{1}, (0, 0, 0, 0, -1, 0)$ & $\mathbbm{1}, (0, 0, 0, 0, 0, 0)$ & $\mathbbm{1}, (0, 0, 0, 0, 0, 0)$\\
$\theta, (0, 0, 0, 0, 1, 1)$ & $\mathbbm{1}, (0, 0, 0, 0, 0, 0) $ & $\mathbbm{1}, (0, 0, 0, 0, -1, -1)$ & $\mathbbm{1}, (0, 0, 0, 0, 0, 0)$ & $\mathbbm{1}, (0, 0, 0, 0, 0, 0)$\\
$\theta, (0, 0, 1, 0, 1, 0)$ & $\mathbbm{1}, (0, 0, -1, 0, 0, 0) $ & $\mathbbm{1}, (0, 0, 0, 0, -1, 0)$ & $\mathbbm{1}, (0, 0, 0, 0, 0, 0)$ & $\mathbbm{1}, (0, 0, 0, 0, 0, 0)$\\
$\theta, (0, 0, 1, 0, 1, 1)$ & $\mathbbm{1}, (0, 0, -1, 0, 0, 0) $ & $\mathbbm{1}, (0, 0, 0, 0, -1, -1)$ & $\mathbbm{1}, (0, 0, 0, 0, 0, 0)$ & $\mathbbm{1}, (0, 0, 0, 0, 0, 0)$\\
$\theta, (1, 0, 0, 0, 1, 0)$ & $\mathbbm{1}, (-1, 0, 0, 0, 0, 0) $ & $\mathbbm{1}, (0, 0, 0, 0, -1, 0)$ & $\mathbbm{1}, (-1, -1, 0, 0, 0, 0)$ & $\mathbbm{1}, (-1, 0, 0, 0, 0, 0)$\\
$\theta, (1, 0, 0, 0, 1, 1)$ & $\mathbbm{1}, (-1, 0, 0, 0, 0, 0) $ & $\mathbbm{1}, (0, 0, 0, 0, -1, -1)$ & $\mathbbm{1}, (-1, -1, 0, 0, 0, 0)$ & $\mathbbm{1}, (-1, 0, 0, 0, 0, 0)$\\
$\theta, (1, 0, 1, 0, 1, 0)$ & $\mathbbm{1}, (-1, 0, -1, 0, 0, 0) $ & $\mathbbm{1}, (0, 0, 0, 0, -1, 0)$ & $\mathbbm{1}, (-1, -1, 0, 0, 0, 0)$ & $\mathbbm{1}, (-1, 0, 0, 0, 0, 0)$\\
$\theta, (1, 0, 1, 0, 1, 1)$ & $\mathbbm{1}, (-1, 0, -1, 0, 0, 0) $ & $\mathbbm{1}, (0, 0, 0, 0, -1, -1)$ & $\mathbbm{1}, (-1, -1, 0, 0, 0, 0)$ & $\mathbbm{1}, (-1, 0, 0, 0, 0, 0)$\\
$\theta^2, (0, 0, 0, 0, 0, 0)$ & $\mathbbm{1}, (0, 0, 0, 0, 0, 0) $ & $\mathbbm{1}, (0, 0, 0, 0, 0, 0)$ & $\mathbbm{1}, (0, 0, 0, 0, 0, 0)$ & $\mathbbm{1}, (0, 0, 0, 0, 0, 0)$\\
$\theta^2, (0, 0, 0, 1, 0, 0)$ & $\theta, (0, 0, 0, 0, 0, 0) $ & $\mathbbm{1}, (0, 0, 0, 0, 0, 0)$ & $\mathbbm{1}, (0, 0, 0, 0, 0, 0)$ & $\mathbbm{1}, (0, 0, 0, 0, 0, 0)$\\
$\theta^2, (0, 0, 1, 1, 0, 0)$ & $\mathbbm{1}, (0, 0, -1, 0, 0, 0) $ & $\mathbbm{1}, (0, 0, 0, 0, 0, 0)$ & $\mathbbm{1}, (0, 0, 0, 0, 0, 0)$ & $\mathbbm{1}, (0, 0, 0, 0, 0, 0)$\\
$\theta^2, (0, 1, 0, 0, 0, 0)$ & $\theta, (0, 0, 0, 0, 0, 0) $ & $\mathbbm{1}, (0, 0, 0, 0, 0, 0)$ & $\mathbbm{1}, (0, -1, 0, 0, 0, 0)$ & $\theta, (0, 0, 0, 0, 0, 0)$\\
$\theta^2, (0, 1, 0, 1, 0, 0)$ & $\theta, (0, 0, 0, 0, 0, 0) $ & $\mathbbm{1}, (0, 0, 0, 0, 0, 0)$ & $\mathbbm{1}, (0, -1, 0, 0, 0, 0)$ & $\theta, (0, 0, 0, 0, 0, 0)^\dagger$\\
$\theta^2, (0, 1, 1, 0, 0, 0)$ & $\theta, (0, 0, 0, 0, 0, 0) $ & $\mathbbm{1}, (0, 0, 0, 0, 0, 0)$ & $\mathbbm{1}, (0, -1, 0, 0, 0, 0)$ & $\theta, (0, 0, 1, 0, 0, 0)^\dagger$\\
$\theta^2, (0, 1, 1, 1, 0, 0)$ & $\theta, (0, 0, 0, 0, 0, 0) $ & $\mathbbm{1}, (0, 0, 0, 0, 0, 0)$ & $\mathbbm{1}, (0, -1, 0, 0, 0, 0)$ & $\theta, (0, 0, 1, 0, 0, 0)$\\
$\theta^2, (1, 1, 0, 0, 0, 0)$ & $\mathbbm{1}, (-1, 0, 0, 0, 0, 0) $ & $\mathbbm{1}, (0, 0, 0, 0, 0, 0)$ & $\mathbbm{1}, (-1, -1, 0, 0, 0, 0)$ & $\mathbbm{1}, (-1, 0, 0, 0, 0, 0)$\\
$\theta^2, (1, 1, 0, 1, 0, 0)$ & $\theta, (0, 0, 0, 0, 0, 0) $ & $\mathbbm{1}, (0, 0, 0, 0, 0, 0)$ & $\mathbbm{1}, (-1, -1, 0, 0, 0, 0)$ & $\mathbbm{1}, (-1, 0, 0, 0, 0, 0)$\\
$\theta^2, (1, 1, 1, 1, 0, 0)$ & $\mathbbm{1}, (-1, 0, -1, 0, 0, 0) $ & $\mathbbm{1}, (0, 0, 0, 0, 0, 0)$ & $\mathbbm{1}, (-1, -1, 0, 0, 0, 0)$ & $\mathbbm{1}, (-1, 0, 0, 0, 0, 0)$\\
$\theta^3, (0, 0, 0, 0, 0, 0)$ & $\mathbbm{1}, (0, 0, 0, 0, 0, 0) $ & $\mathbbm{1}, (0, 0, 0, 0, 0, 0)$ & $\mathbbm{1}, (0, 0, 0, 0, 0, 0)$ & $\mathbbm{1}, (0, 0, 0, 0, 0, 0)$\\
$\theta^3, (0, 0, 0, 0, 0, 1)$ & $\mathbbm{1}, (0, 0, 0, 0, 0, 0) $ & $\mathbbm{1}, (0, 0, 0, 0, 0, -1)$ & $\mathbbm{1}, (0, 0, 0, 0, 0, 0)$ & $\mathbbm{1}, (0, 0, 0, 0, 0, 0)$\\
$\theta^3, (0, 0, 1, 0, 0, 0)$ & $\mathbbm{1}, (0, 0, 0, 1, 0, 0) $ & $\mathbbm{1}, (0, 0, 0, 0, 0, 0)$ & $\mathbbm{1}, (0, 0, 0, 0, 0, 0)$ & $\mathbbm{1}, (0, 0, 0, 0, 0, 0)$\\
$\theta^3, (0, 0, 1, 0, 0, 1)$ & $\mathbbm{1}, (0, 0, 0, 1, 0, 0) $ & $\mathbbm{1}, (0, 0, 0, 0, 0, -1)$ & $\mathbbm{1}, (0, 0, 0, 0, 0, 0)$ & $\mathbbm{1}, (0, 0, 0, 0, 0, 0)$\\
$\theta^3, (1, 0, 0, 0, 0, 0)$ & $\mathbbm{1}, (0, 1, 0, 0, 0, 0) $ & $\mathbbm{1}, (0, 0, 0, 0, 0, 0)$ & $\mathbbm{1}, (-1, 1, 0, 0, 0, 0)$ & $\mathbbm{1}, (0, 1, 0, 0, 0, 0)$\\
$\theta^3, (1, 0, 0, 0, 0, 1)$ & $\mathbbm{1}, (0, 1, 0, 0, 0, 0) $ & $\mathbbm{1}, (0, 0, 0, 0, 0, -1)$ & $\mathbbm{1}, (-1, 1, 0, 0, 0, 0)$ & $\mathbbm{1}, (0, 1, 0, 0, 0, 0)$\\
$\theta^3, (1, 0, 1, 0, 0, 0)$ & $\mathbbm{1}, (0, 1, 0, 1, 0, 0) $ & $\mathbbm{1}, (0, 0, 0, 0, 0, 0)$ & $\mathbbm{1}, (-1, 1, 0, 0, 0, 0)$ & $\mathbbm{1}, (0, 1, 0, 0, 0, 0)$\\
$\theta^3, (1, 0, 1, 0, 0, 1)$ & $\mathbbm{1}, (0, 1, 0, 1, 0, 0) $ & $\mathbbm{1}, (0, 0, 0, 0, 0, -1)$ & $\mathbbm{1}, (-1, 1, 0, 0, 0, 0)$ & $\mathbbm{1}, (0, 1, 0, 0, 0, 0)$\\
$\theta^3, (0, 0, 0, 0, 1, 0)$ & $\mathbbm{1}, (0, 0, 0, 0, 0, 0) $ & $\mathbbm{1}, (0, 0, 0, 0, -1, 0)$ & $\mathbbm{1}, (0, 0, 0, 0, 0, 0)$ & $\mathbbm{1}, (0, 0, 0, 0, 0, 0)$\\
$\theta^3, (0, 0, 0, 0, 1, 1)$ & $\mathbbm{1}, (0, 0, 0, 0, 0, 0) $ & $\mathbbm{1}, (0, 0, 0, 0, -1, -1)$ & $\mathbbm{1}, (0, 0, 0, 0, 0, 0)$ & $\mathbbm{1}, (0, 0, 0, 0, 0, 0)$\\
$\theta^3, (0, 0, 1, 0, 1, 0)$ & $\mathbbm{1}, (0, 0, 0, 1, 0, 0) $ & $\mathbbm{1}, (0, 0, 0, 0, -1, 0)$ & $\mathbbm{1}, (0, 0, 0, 0, 0, 0)$ & $\mathbbm{1}, (0, 0, 0, 0, 0, 0)$\\
$\theta^3, (0, 0, 1, 0, 1, 1)$ & $\mathbbm{1}, (0, 0, 0, 1, 0, 0) $ & $\mathbbm{1}, (0, 0, 0, 0, -1, -1)$ & $\mathbbm{1}, (0, 0, 0, 0, 0, 0)$ & $\mathbbm{1}, (0, 0, 0, 0, 0, 0)$\\
$\theta^3, (1, 0, 0, 0, 1, 0)$ & $\mathbbm{1}, (0, 1, 0, 0, 0, 0) $ & $\mathbbm{1}, (0, 0, 0, 0, -1, 0)$ & $\mathbbm{1}, (-1, 1, 0, 0, 0, 0)$ & $\mathbbm{1}, (0, 1, 0, 0, 0, 0)$\\
$\theta^3, (1, 0, 0, 0, 1, 1)$ & $\mathbbm{1}, (0, 1, 0, 0, 0, 0) $ & $\mathbbm{1}, (0, 0, 0, 0, -1, -1)$ & $\mathbbm{1}, (-1, 1, 0, 0, 0, 0)$ & $\mathbbm{1}, (0, 1, 0, 0, 0, 0)$\\
$\theta^3, (1, 0, 1, 0, 1, 0)$ & $\mathbbm{1}, (0, 1, 0, 1, 0, 0) $ & $\mathbbm{1}, (0, 0, 0, 0, -1, 0)$ & $\mathbbm{1}, (-1, 1, 0, 0, 0, 0)$ & $\mathbbm{1}, (0, 1, 0, 0, 0, 0)$\\
$\theta^3, (1, 0, 1, 0, 1, 1)$ & $\mathbbm{1}, (0, 1, 0, 1, 0, 0) $ & $\mathbbm{1}, (0, 0, 0, 0, -1, -1)$ & $\mathbbm{1}, (-1, 1, 0, 0, 0, 0)$ & $\mathbbm{1}, (0, 1, 0, 0, 0, 0)$\\
\end{tabular}
\caption{Values for $h_g$'s for $\mathbbm{Z}_4$. The elements marked with $^\dagger$ correspond to the $h_g$ and $h_{g^\prime}$ from eq. \protect\eqref{ExampleHgs}.}\label{S:AppHgZ4}
\end{table}

\begin{table}[h]
\centering
\small
\begin{tabular}{  c | c  c  c  c }
$g$ & $h_g^{\theta_1}$ & $h_g^{\theta_2}$ & $h_g^{\theta_3}$\\
\hline
$\mathbbm{1}, (0, 0, 0, 0, 0, 0)$ & $\mathbbm{1}, (0, 0, 0, 0, 0, 0) $ & $\mathbbm{1}, (0, 0, 0, 0, 0, 0)$ & $\mathbbm{1}, (0, 0, 0, 0, 0, 0)$ \\
$\theta, (0, 0, 1, 1, 1, 1)$ & $\mathbbm{1}, (0, 0, 0, 0, 0, 0) $ & $\mathbbm{1}, (0, 0, -1, -1, 0, 0)$ & $\mathbbm{1}, (0, 0, 0, 0, -1, -1)$ \\
$\theta, (0, 0, 1, 1, 1, 0)$ & $\mathbbm{1}, (0, 0, 0, 0, 0, 0) $ & $\mathbbm{1}, (0, 0, -1, -1, 0, 0)$ & $\mathbbm{1}, (0, 0, 0, 0, -1, 0)$ \\
$\theta, (0, 0, 1, 1, 0, 1)$ & $\mathbbm{1}, (0, 0, 0, 0, 0, 0) $ & $\mathbbm{1}, (0, 0, -1, -1, 0, 0)$ & $\mathbbm{1}, (0, 0, 0, 0, 0, -1)$ \\
$\theta, (0, 0, 1, 1, 0, 0)$ & $\mathbbm{1}, (0, 0, 0, 0, 0, 0) $ & $\mathbbm{1}, (0, 0, -1, -1, 0, 0)$ & $\mathbbm{1}, (0, 0, 0, 0, 0, 0)$ \\
$\theta, (0, 0, 0, 0, 1, 1)$ & $\mathbbm{1}, (0, 0, 0, 0, 0, 0) $ & $\mathbbm{1}, (0, 0, 0, 0, 0, 0)$ & $\mathbbm{1}, (0, 0, 0, 0, -1, -1)$ \\
$\theta, (0, 0, 0, 0, 1, 0)$ & $\mathbbm{1}, (0, 0, 0, 0, 0, 0) $ & $\mathbbm{1}, (0, 0, 0, 0, 0, 0)$ & $\mathbbm{1}, (0, 0, 0, 0, -1, 0)$ \\
$\theta, (0, 0, 0, 0, 0, 1)$ & $\mathbbm{1}, (0, 0, 0, 0, 0, 0) $ & $\mathbbm{1}, (0, 0, 0, 0, 0, 0)$ & $\mathbbm{1}, (0, 0, 0, 0, 0, -1)$ \\
$\theta, (0, 0, 0, 0, 0, 0)$ & $\mathbbm{1}, (0, 0, 0, 0, 0, 0) $ & $\mathbbm{1}, (0, 0, 0, 0, 0, 0)$ & $\mathbbm{1}, (0, 0, 0, 0, 0, 0)$ \\
$\theta, (0, 0, 1, 0, 1, 1)$ & $\mathbbm{1}, (0, 0, 0, 0, 0, 0) $ & $\mathbbm{1}, (0, 0, -1, 0, 0, 0)$ & $\mathbbm{1}, (0, 0, 0, 0, -1, -1)$ \\
$\theta, (0, 0, 1, 0, 1, 0)$ & $\mathbbm{1}, (0, 0, 0, 0, 0, 0) $ & $\mathbbm{1}, (0, 0, -1, 0, 0, 0)$ & $\mathbbm{1}, (0, 0, 0, 0, -1, 0)$ \\
$\theta, (0, 0, 1, 0, 0, 1)$ & $\mathbbm{1}, (0, 0, 0, 0, 0, 0) $ & $\mathbbm{1}, (0, 0, -1, 0, 0, 0)$ & $\mathbbm{1}, (0, 0, 0, 0, 0, -1)$ \\
$\theta, (0, 0, 1, 0, 0, 0)$ & $\mathbbm{1}, (0, 0, 0, 0, 0, 0) $ & $\mathbbm{1}, (0, 0, -1, 0, 0, 0)$ & $\mathbbm{1}, (0, 0, 0, 0, 0, 0)$ \\
$\theta^2, (-1, 1, 0, 2, 0, 0)$ & $\theta, (0, 0, 2, 2, 0, 0) $ & $\mathbbm{1}, (0, 0, -2, -2, 0, 0)$ & $\mathbbm{1}, (0, 0, 0, 0, 0, 0)$ \\
$\theta^2, (-1, 1, 0, 0, 0, 0)$ & $\theta, (0, 0, 0, 0, 0, 0) $ & $\mathbbm{1}, (0, 0, 0, 0, 0, 0)$ & $\mathbbm{1}, (0, 0, 0, 0, 0, 0)$ \\
$\theta^2, (-1, 1, 0, 1, 0, 0)$ & $\theta, (0, 0, 1, 1, 0, 0) $ & $\mathbbm{1}, (0, 0, -1, -1, 0, 0)$ & $\mathbbm{1}, (0, 0, 0, 0, 0, 0)$ \\
$\theta^2, (0, 0, 0, 2, 0, 0)$ & $\mathbbm{1}, (0, 0, 0, 0, 0, 0) $ & $\mathbbm{1}, (0, 0, -2, -2, 0, 0)$ & $\mathbbm{1}, (0, 0, 0, 0, 0, 0)$ \\
$\theta^2, (0, 0, 0, 0, 0, 0)$ & $\mathbbm{1}, (0, 0, 0, 0, 0, 0) $ & $\mathbbm{1}, (0, 0, 0, 0, 0, 0)$ & $\mathbbm{1}, (0, 0, 0, 0, 0, 0)$ \\
$\theta^2, (0, 0, 0, 1, 0, 0)$ & $\mathbbm{1}, (0, 0, 0, 0, 0, 0) $ & $\mathbbm{1}, (0, 0, -1, -1, 0, 0)$ & $\mathbbm{1}, (0, 0, 0, 0, 0, 0)$ \\
$\theta^3, (1, 0, 0, 0, 1, 1)$ & $\theta, (0, 0, 0, 0, 1, 1) $ & $\mathbbm{1}, (0, 0, 0, 0, 0, 0)$ & $\mathbbm{1}, (0, 0, 0, 0, -1, -1)$ \\
$\theta^3, (1, 0, 0, 0, 1, 0)$ & $\theta, (0, 0, 0, 0, 1, 0) $ & $\mathbbm{1}, (0, 0, 0, 0, 0, 0)$ & $\mathbbm{1}, (0, 0, 0, 0, -1, 0)$ \\
$\theta^3, (1, 0, 0, 0, 0, 1)$ & $\theta, (0, 0, 0, 0, 0, 1) $ & $\mathbbm{1}, (0, 0, 0, 0, 0, 0)$ & $\mathbbm{1}, (0, 0, 0, 0, 0, -1)$ \\
$\theta^3, (1, 0, 0, 0, 0, 0)$ & $\theta, (0, 0, 0, 0, 0, 0) $ & $\mathbbm{1}, (0, 0, 0, 0, 0, 0)$ & $\mathbbm{1}, (0, 0, 0, 0, 0, 0)$ \\
$\theta^3, (0, 0, 0, 0, 1, 1)$ & $\mathbbm{1}, (0, 0, 0, 0, 0, 0) $ & $\mathbbm{1}, (0, 0, 0, 0, 0, 0)$ & $\mathbbm{1}, (0, 0, 0, 0, -1, -1)$ \\
$\theta^3, (0, 0, 0, 0, 1, 0)$ & $\mathbbm{1}, (0, 0, 0, 0, 0, 0) $ & $\mathbbm{1}, (0, 0, 0, 0, 0, 0)$ & $\mathbbm{1}, (0, 0, 0, 0, -1, 0)$ \\
$\theta^3, (0, 0, 0, 0, 0, 1)$ & $\mathbbm{1}, (0, 0, 0, 0, 0, 0) $ & $\mathbbm{1}, (0, 0, 0, 0, 0, 0)$ & $\mathbbm{1}, (0, 0, 0, 0, 0, -1)$ \\
$\theta^3, (0, 0, 0, 0, 0, 0)$ & $\mathbbm{1}, (0, 0, 0, 0, 0, 0) $ & $\mathbbm{1}, (0, 0, 0, 0, 0, 0)$ & $\mathbbm{1}, (0, 0, 0, 0, 0, 0)$ \\
$\theta^4, (-1, 1, 1, 1, 0, 0)$ & $\theta, (0, 0, 1, 1, 0, 0) $ & $\mathbbm{1}, (0, 0, -1, -1, 0, 0)$ & $\mathbbm{1}, (0, 0, 0, 0, 0, 0)$ \\
$\theta^4, (-1, 1, 0, 0, 0, 0)$ & $\theta, (0, 0, 0, 0, 0, 0) $ & $\mathbbm{1}, (0, 0, 0, 0, 0, 0)$ & $\mathbbm{1}, (0, 0, 0, 0, 0, 0)$ \\
$\theta^4, (-1, 1, 1, 0, 0, 0)$ & $\theta, (0, 0, 1, 0, 0, 0) $ & $\mathbbm{1}, (0, 0, -1, 0, 0, 0)$ & $\mathbbm{1}, (0, 0, 0, 0, 0, 0)$ \\
$\theta^4, (0, 0, 1, 1, 0, 0)$ & $\mathbbm{1}, (0, 0, 0, 0, 0, 0) $ & $\mathbbm{1}, (0, 0, -1, -1, 0, 0)$ & $\mathbbm{1}, (0, 0, 0, 0, 0, 0)$ \\
$\theta^4, (0, 0, 0, 0, 0, 0)$ & $\mathbbm{1}, (0, 0, 0, 0, 0, 0) $ & $\mathbbm{1}, (0, 0, 0, 0, 0, 0)$ & $\mathbbm{1}, (0, 0, 0, 0, 0, 0)$ \\
$\theta^4, (0, 0, 1, 0, 0, 0)$ & $\mathbbm{1}, (0, 0, 0, 0, 0, 0) $ & $\mathbbm{1}, (0, 0, -1, 0, 0, 0)$ & $\mathbbm{1}, (0, 0, 0, 0, 0, 0)$ \\
$\theta^5, (0, 0, 0, 2, 1, 1)$ & $\mathbbm{1}, (0, 0, 0, 0, 0, 0) $ & $\mathbbm{1}, (0, 0, -2, -2, 0, 0)$ & $\mathbbm{1}, (0, 0, 0, 0, -1, -1)$ \\
$\theta^5, (0, 0, 0, 2, 1, 0)$ & $\mathbbm{1}, (0, 0, 0, 0, 0, 0) $ & $\mathbbm{1}, (0, 0, -2, -2, 0, 0)$ & $\mathbbm{1}, (0, 0, 0, 0, -1, 0)$ \\
$\theta^5, (0, 0, 0, 2, 0, 1)$ & $\mathbbm{1}, (0, 0, 0, 0, 0, 0) $ & $\mathbbm{1}, (0, 0, -2, -2, 0, 0)$ & $\mathbbm{1}, (0, 0, 0, 0, 0, -1)$ \\
$\theta^5, (0, 0, 0, 2, 0, 0)$ & $\mathbbm{1}, (0, 0, 0, 0, 0, 0) $ & $\mathbbm{1}, (0, 0, -2, -2, 0, 0)$ & $\mathbbm{1}, (0, 0, 0, 0, 0, 0)$ \\
$\theta^5, (0, 0, 0, 0, 1, 1)$ & $\mathbbm{1}, (0, 0, 0, 0, 0, 0) $ & $\mathbbm{1}, (0, 0, 0, 0, 0, 0)$ & $\mathbbm{1}, (0, 0, 0, 0, -1, -1)$ \\
$\theta^5, (0, 0, 0, 0, 1, 0)$ & $\mathbbm{1}, (0, 0, 0, 0, 0, 0) $ & $\mathbbm{1}, (0, 0, 0, 0, 0, 0)$ & $\mathbbm{1}, (0, 0, 0, 0, -1, 0)$ \\
$\theta^5, (0, 0, 0, 0, 0, 1)$ & $\mathbbm{1}, (0, 0, 0, 0, 0, 0) $ & $\mathbbm{1}, (0, 0, 0, 0, 0, 0)$ & $\mathbbm{1}, (0, 0, 0, 0, 0, -1)$ \\
$\theta^5, (0, 0, 0, 0, 0, 0)$ & $\mathbbm{1}, (0, 0, 0, 0, 0, 0) $ & $\mathbbm{1}, (0, 0, 0, 0, 0, 0)$ & $\mathbbm{1}, (0, 0, 0, 0, 0, 0)$ \\
$\theta^5, (0, 0, 0, 1, 1, 1)$ & $\mathbbm{1}, (0, 0, 0, 0, 0, 0) $ & $\mathbbm{1}, (0, 0, -1, -1, 0, 0)$ & $\mathbbm{1}, (0, 0, 0, 0, -1, -1)$ \\
$\theta^5, (0, 0, 0, 1, 1, 0)$ & $\mathbbm{1}, (0, 0, 0, 0, 0, 0) $ & $\mathbbm{1}, (0, 0, -1, -1, 0, 0)$ & $\mathbbm{1}, (0, 0, 0, 0, -1, 0)$ \\
$\theta^5, (0, 0, 0, 1, 0, 1)$ & $\mathbbm{1}, (0, 0, 0, 0, 0, 0) $ & $\mathbbm{1}, (0, 0, -1, -1, 0, 0)$ & $\mathbbm{1}, (0, 0, 0, 0, 0, -1)$ \\
$\theta^5, (0, 0, 0, 1, 0, 0)$ & $\mathbbm{1}, (0, 0, 0, 0, 0, 0) $ & $\mathbbm{1}, (0, 0, -1, -1, 0, 0)$ & $\mathbbm{1}, (0, 0, 0, 0, 0, 0)$ \\
\end{tabular}
\caption{Values for $h_g$'s for $\mathbbm{Z}_{6-II}$.}\label{S:AppHgZ6II}
\end{table}
\end{appendix}

\end{document}